# 1D-confined crystallization routes for tungsten phosphides

*Gangtae Jin,[1,5] Christian D. Multunas, [2,5] James L. Hart, [1,5] Mehrdad T. Kiani,[1] Quynh P. Sam,[1] Han Wang,[1] Yeryun Cheon,[1] Khoan Duong,[1] David J. Hynek, [3] Hyeuk Jin Han,[4*] Ravishankar Sundararaman [2*], and Judy J. Cha[1*]*

[1]Department of Materials Science and Engineering, Cornell University, Ithaca, New York 14850, USA

[2] Department of Materials Science and Engineering, Rensselaer Polytechnic Institute, Troy, NY, 12180 USA

[3] Department of Mechanical Engineering and Materials Science, Yale University, New Haven, Connecticut 06511, USA

[4] Department of Environment and Energy Engineering, Sungshin Women's University, Seoul 01133, Republic of Korea

[5]These authors contributed equally: Gangtae Jin, Christian D. Multunas, James L. Hart.

[*]Corresponding authors:

E-mail: judy.cha@cornell.edu, sundar@rpi.edu, hyeukjin.han@sungshin.ac.kr

**ABSTRACT:**

Topological materials confined in one-dimension (1D) can transform computing technologies, such as 1D topological semimetals for nanoscale interconnects and 1D topological superconductors for fault-tolerant quantum computing. As such, understanding crystallization of 1D-confined topological materials is critical. Here, we demonstrate 1D-confined crystallization routes during template-assisted nanowire synthesis where we observe diameter-dependent phase selectivity for topological metal tungsten phosphides. A phase bifurcation occurs to produce tungsten monophosphide and tungsten diphosphide at the cross-over nanowire diameter of ~ 35 nm. Four-dimensional scanning transmission electron microscopy was used to identify the two phases and to map crystallographic orientations of grains at a few nm resolution. The 1D-confined phase selectivity is attributed to the minimization of the total surface energy, which depends on the nanowire diameter and chemical potentials of precursors. Theoretical calculations were carried out to construct the diameter-dependent phase diagram, which agrees with experimental observations. Our findings suggest a new crystallization route to stabilize topological materials confined in 1D.

**Introduction**

Nanostructured transition metal phosphides (TMPs) are a promising material platform for energy storage, catalysis, and photonics.[1-6] A subset of TMPs possess topologically non-trivial electronic band structures: group V phosphides such as niobium phosphide and tantalum phosphide are Weyl semimetals,[7-9] and group VI phosphides such as molybdenum monophosphide (MoP) and tungsten monophosphide (WP) are topological metals that exhibit high conductivity and high carrier density, along with the topologically protected fermions.[10-13] Additional interesting properties for WP include superconductivity with a small electron-phonon coupling strength[14-17] and multiple semi-Dirac-like points near the Fermi level.[17] Tungsten diphosphide ($WP_2$) is another topological TMP that exhibits high magnetoresistance ($3 \times 10^5$ %) and low resistivity values of 3–4 nΩ cm at 2 K, which was attributed to the robust topological phase with two neighboring Weyl points.[18-20] These TMP topological semimetals[8-20] are an emerging class of nanoscale interconnect materials that can potentially deliver the desired dimensional scaling of decreasing resistivity with decreasing dimensions, arising from the topological surface states and suppressed electron backscattering.[21-24] Despite the promises for low-resistance nanoscaled interconnects, controlled synthesis of nanostructured topological semimetals has been under-investigated. Nevertheless, precision synthesis of 1D-confined TMPs is essential for the realization of energy-efficient computing technologies based on their emergent transport phenomena[25-28].

For crystallization in 1D, the free energy landscape, which underlies the kinetics and thermodynamics of the crystallization process, is significantly affected by the nanoscale confinement.[29-31] At large surface-to-volume ratios, surface energy of the growth products can dominate the crystallization pathway: a metastable phase might be preferred over a stable phase if the surface energy of the metastable phase is lower. Thus, nanoscale confinement can be exploited to control phase stability and synthesis pathways.[32,33] Here, we developed a 1D-confined synthesis of topological metal WP and $WP_2$ via 1D template-assisted transformations. With decreasing diameter, we observe unexpected crystallization pathways, resulting in distinct phases. The size-effects on phase stability of tungsten phosphides were studied using four-dimensional scanning transmission electron microscopy (4D-STEM) and automated crystal orientation mapping (ACOM) on 4D STEM datasets.[34,35] Diameter-dependent phase selectivity of WP or WP$_2$ is attributed to the surface energy differences of the synthesized nanowires. A diameter-dependent phase diagram was constructed by density-functional-theory (DFT) calculations, which support 1D-confined crystallization routes between WP and α-WP$_2$. For the interconnect applications, we measured resistivity of the synthesized nanowires and show that polycrystalline 1D-WP should have minimized surface electron scatterings due to its small mean free path of 3.15 nm. Our findings demonstrate that diameter-dependent crystallization is a viable synthesis route for 1D topological semimetals.

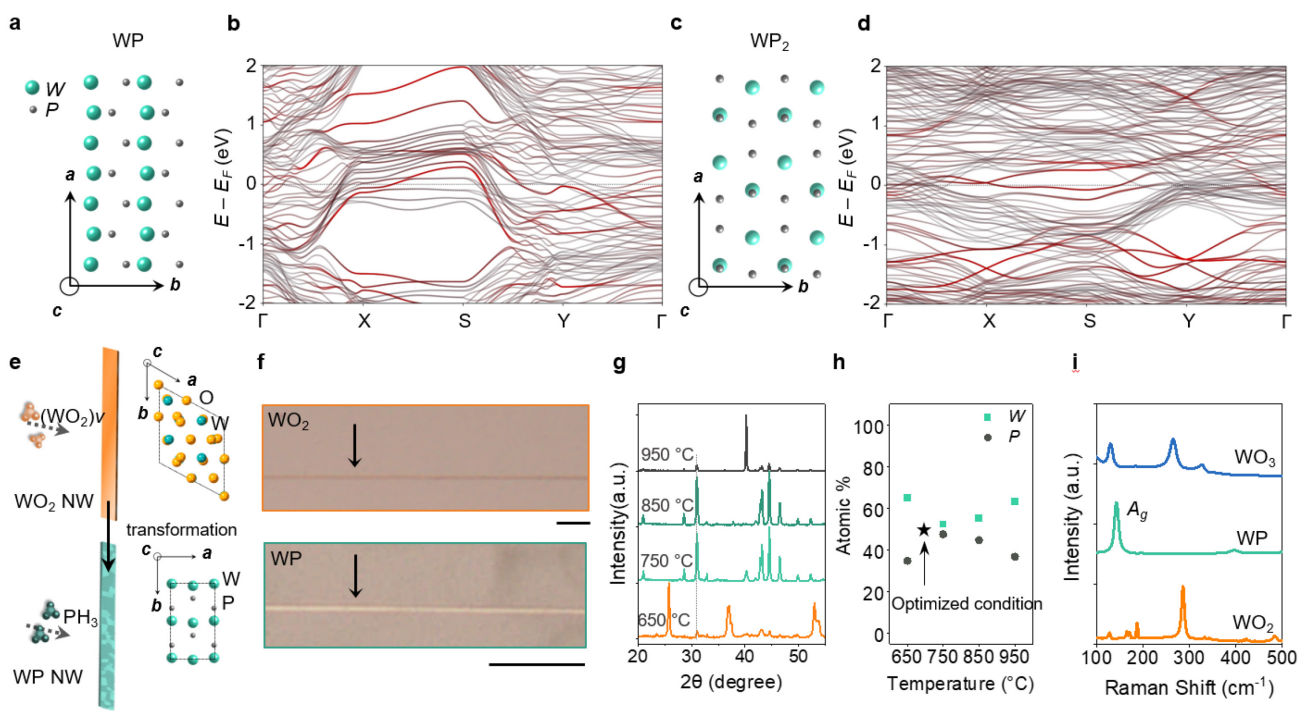

**Fig. 1| Electron band structures and template-assisted conversion of WP and WP$_2$.**
**a,b,** Crystal structure (**a**) and calculated surface-weighted electronic band structure (**b**) of a WP slab, terminated with the lowest surface energy plane (011). Surface states from the terminated crystal planes are colored in red. Note that these surface states do not represent topologically protected surface states. **c,d,** Crystal structure (**c**) and calculated surface-weighted electronic band structure (**d**) of a α-WP$_2$ slab, terminated with the lowest surface energy plane (110). Surface states from the terminated crystal planes are colored in red. We have chosen the same high-symmetry path for orthorhombic WP and monoclinic $WP_2$ since these band structures represent slabs rather than bulk structures, and the slabs have identical surface unit cells. **e**, Schematics of geometrically confined transformation from $WO_2$ to WP with corresponding crystal structures of $WO_2$ and WP. **f**, Optical microscope images of $WO_2$ template grown on c-cut sapphire and transformed 1D-WP. scale bars: 10 μm. **g**, Powder X-ray diffraction spectra of growth products as a function of the conversion temperature. **h**, Relative atomic content of $W_xP_{1-x}$ (0.5<x<0.7) growth products, collected from SEM-EDS signals at 4 different conversion temperatures. **i**, Raman spectra of as grown 1D-$WO_2$ (orange), transformed 1D-WP (green), and $WO_3$ (blue), respectively. a.u., arbitrary units.

## Results and Discussion

### Band structure and synthesis of 1D tungsten phosphide by template-assisted transformation

DFT calculations were performed to obtain electronic band structures of WP and $WP_2$. WP has an orthorombic crystal structure (Fig. 1a) with lattice parameters a = 0.327 nm, b = 0.576 nm, c = 0.627 nm.[17] To represent nanoscale WP with large surface-to-volume ratios, we calculated a surface-weighted band structure of an 8-unit cell thick WP slab that was terminated with the lowest surface energy crystal planes of (011) (Fig. 1b). Monoclinic α-$WP_2$ with lattice parameters a = 0.848 nm, b = 0.319 nm, c = 0.748 nm is one of the stable W-P compounds at room temperature (Fig. 1c). α-$WP_2$ is topologically trivial unlike β-$WP_2$ that has Weyl nodes.[36] The surface-weighted band structure of a monoclinic 11-unit cell thick α-$WP_2$ slab terminated with (110) crystal planes is shown in Fig. 1d. Band structure calculations show both are metals.

Highly anisotropic WP nanostructures were synthesized via 1D-confined transformation by converting $WO_2$ nanowires to WP nanowires, as illustrated in Fig. 1e. First, we grew 1D-$WO_2$ on the *c*-sapphire substrates with a miscut angle of 1° along the A-axis (⟨11–20⟩)[37,38] by vapor transport synthesis from $WO_3$ powder precursors. As-grown 1D-$WO_2$ templates (Monoclinic, $P2_1/c$ space group, a = 0.576 nm, b = 0.484 nm, c = 0.580 nm) were transformed to orthorhombic WP via phosphorization using $PH_3$ gas, produced from the thermal decomposition of $NaH_2PO_2·H_2O$ at 700 °C (see Methods and Supplementary Fig. 1 for growth details). The transformed WP exhibit highly directional crystalline form with typical widths of 100 ~ 500 nm and lengths of 10 ~ 20 μm as shown in Fig. 1f. The conversion from $WO_2$ to WP is accompanied by the volume decrease from the unit cell volume of 149.31 Å$^3$ for WO$_2$ to 117.90 Å$^3$ for WP due to the change in crystal symmetry and lattice parameters. The WO$_2$ to WP transformation is apparent by the color change in converted crystals, but the original morphologies of the WO$_2$ nanowires were preserved in converted WP.

The conversion temperature was optimized by analyzing X-ray diffraction (XRD) and Energy dispersive X-ray spectroscopy (EDS) data (Fig. 1g and 1h). At 650 ˚C, the product still contained WO$_2$ diffraction peaks and excess W with an elemental ratio of W:P = 1.86:1, which suggest incomplete conversion to WP. At higher temperatures (> 750 ˚C), XRD patterns mainly indicate the orthorhombic WP without the formation of other W-P phases, such as $\alpha$-WP$_2$, $\beta$-WP$_2$, WP$_3$, or W$_3$P. However, the stoichiometry of converted WP deviated from the expected W:P ratio of 1:1. With increasing conversion temperature, the W:P ratio was 1.1:1 at 750 ˚C, 1.2:1 at 850 ˚C, and 1.7:1 at 950 ˚C. From the XRD and EDS analysis, the conversion temperature was set to 700 ˚C to achieve the W:P ratio of 1:1. To verify the formation of WP and absence of any residual tungsten oxides, we also obtained the Raman spectra (wavelength $\lambda$=532 nm) of initial WO$_2$ templates and final WP nanostructures converted at 700 ˚C (Figure 1i). The Raman spectrum of the WO$_2$ template (orange) is in agreement with previous studies of 1D WO$_2$.[39] The Raman spectrum of the WP nanostructures (light green) shows a strong scattering peak at 142.3 cm$^{-1}$, indicating the $A_g$ vibration mode.[40] At a higher laser intensity (> 300 µW), the $A_g$ peak disappeared and the WP nanostructures were oxidized to WO$_3$, as supported by the presence of the Raman peaks (blue) at 130 cm$^{-1}$, 264 cm$^{-1}$, and 326 cm$^{-1}$. Thus, we confirm the complete transformation to WP from WO$_2$ at the conversion temperature of 700 ˚C based on XRD, EDS, and Raman spectroscopy.

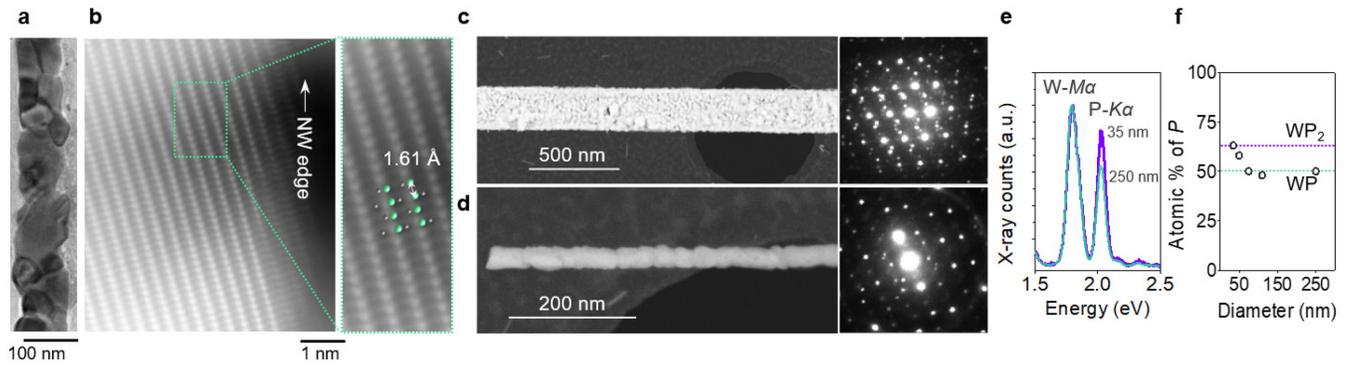

**Fig. 2| Structural characterization of 1D-confined WP$_{x\,(x=1\text{ or }2)}$ with varying diameter.**
**a**, Low-magnification TEM image of 1D-confined WP. **b**, Atomic-resolution HAADF-STEM images of WP obtained from a single grain of the polycrystalline WP. **c,d,** HAADF-STEM images of 1D-confined WP with the diffraction pattern of dominant zone axis of [131] (**c**) and WP$_2$ with the corresponding diffraction pattern of zone axis of [233] (**d**). **e**, TEM-EDS spectra acquired from representative 1D-confined WP$_{x\,(x=1\text{ or }2)}$. **f**, Relative atomic content of P with varying diameter, collected from TEM-EDS signals of 5 different samples.

**Structure characterization of 1D-confined WP with varying diameter**

Using transmission electron microscopy (TEM), we characterized the atomic structure of the 1D-confined WP with various diameters. The TEM image presented in Fig. 2a shows the polycrystalline nature of the 1D-confined WP with nanoscale grains. Individual WP grains are merged to form a closed-packed nanowire (width of ~ 100 nm) without any pores and noticeable oxide layers. The high-angle annular dark-field scanning transmission electron microscopy (HAADF-STEM) image of one of the grains in the nanowire confirms the atomic structure of WP with bright W atomic columns that arise from the atomic number (Z) difference of $Z_P = 15$ and $Z_W = 74$. The lattice image of the transformed WP agrees with the atomic model viewed along the [001] direction of WP (Green: W atoms, Grey: P atoms) (Fig. 2b).

Microstructures of two WP samples with different diameters were examined using HAADF-STEM (250 nm diameter in Fig. 2c and 35 nm diameter in Fig. 2d), which showed increasing porosity and more grains with increasing diameter. Surprisingly, the electron diffraction patterns from the 35 nm- and 250 nm-wide WP nanowires were different, which could not simply be attributed to different tilt angles of the nanowires with respect to the electron beam. The chemical compositions of these two nanowires were analyzed by TEM-EDS (Fig. 2e and 2f), which showed that the 35 ~ 50 nm-wide nanowire had significantly less P by 60 ~ 66 % compared to the 75 ~ 250 nm-wide nanowire, which showed roughly 1:1 atomic ratio of W:P. Several nanowires were analyzed using TEM-EDS and showed that below ~ 50 nm, the W:P ratio started to deviate from 1:1. This suggests the growth products might contain more than one phase of W-P compounds despite our XRD analysis that only showed the monoclinic WP phase (Fig. 1g).

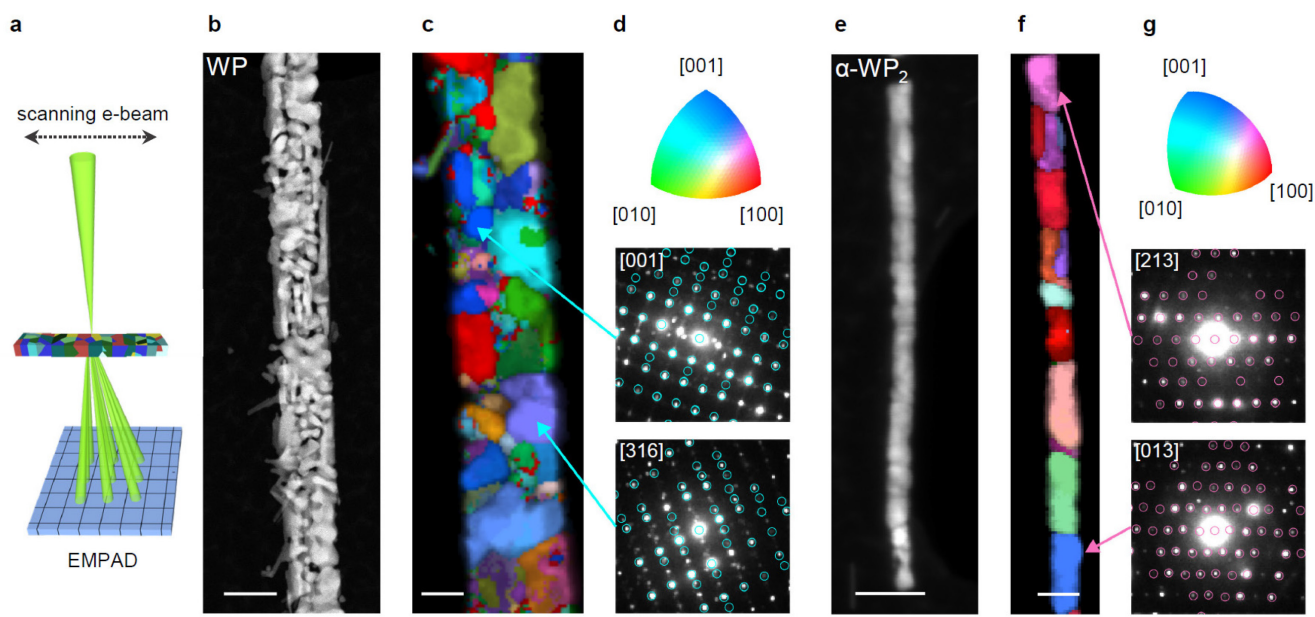

**Fig. 3| 4D STEM and grain orientation mapping of WP and α-WP$_2$.**
**a**, Schematic for 4D-STEM of 1D-confined WP$_x$ $(x = 1$ or $2)$ on EMPAD. **b,c** HAADF-STEM image (**b**) and grain orientation map (**c**) of 1D-confined orthorhombic WP. The orientation map only shows a sub-section of the wire. **d**, Inverse pole figure for WP and electron diffraction patterns with simulated diffraction patterns from specific grains of orthorhombic WP. **e,f** HAADF-STEM image (**e**) and grain orientation map (**f**) of 1D-confined monoclinic WP$_2$. **g**, Inverse pole figure for WP$_2$ and electron diffraction patterns with simulated diffraction patterns from specific grains of monoclinic WP$_2$. Scale bars: 100 nm for **b** and **e**, 50 nm for **c** and **f**. For the overlapping experimental and simulated diffraction patterns, the label provides the nearest zone axis.

**Crystallographic orientation mapping of WP and α-WP$_2$**

To discern additional phases beside WP in the converted phosphides, we carried out 4D-STEM on converted phosphide nanowires to obtain atomic structures of individual grains as illustrated in Fig. 3a (details in Methods). Using the ACOM on the 4D-STEM datasets, nanoscale grain orientation maps were constructed for two poly-crystalline nanowires.[34,35] Figure 3b shows a 100 nm-wide tungsten phosphide, which was identified as orthorhombic WP (space group of Pnma) by comparing the 4D STEM data with simulated diffraction patterns. A nanoscale grain map obtained from the 4D STEM data (Fig. 3c and 3d) suggests that all the grains present in this nanowire are WP with different grain orientations. Note that for this specimen, the domain size is smaller than the nanowire thickness (in the beam direction), such that all the recorded diffraction patterns contain signal from multiple grains. Accordingly, the orientation maps reflect the orientation of the grain which produces the highest diffraction intensity, and the highest correlation with the simulated patterns. In contrast, a 35-nm-wide tungsten phosphide nanowire shown in Fig. 3e was identified to be monoclinic α-$WP_2$ (space group of C2/m) according to the 4D STEM data and subsequent analysis. This can explain the deviation in the W:P ratio of the nanowires with diameter below ∼ 50 nm analyzed by TEM-EDS (Fig. 2f), which showed a W:P ratio of approximately 1:2. The grain map of the α-$WP_2$ nanowire (Fig. 3f and 3g) suggests that all the grains are α-$WP_2$. Thus, using 4D STEM, we observed diameter-dependent phase bifurcation in the crystallization of 1D-confined tungsten phsphides where $WO_2$ is converted to WP or $WP_2$ above and below ∼ 35 nm diameter, respectively. These two phases for the different diameter nanowires were also confirmed with TEM-EDS.

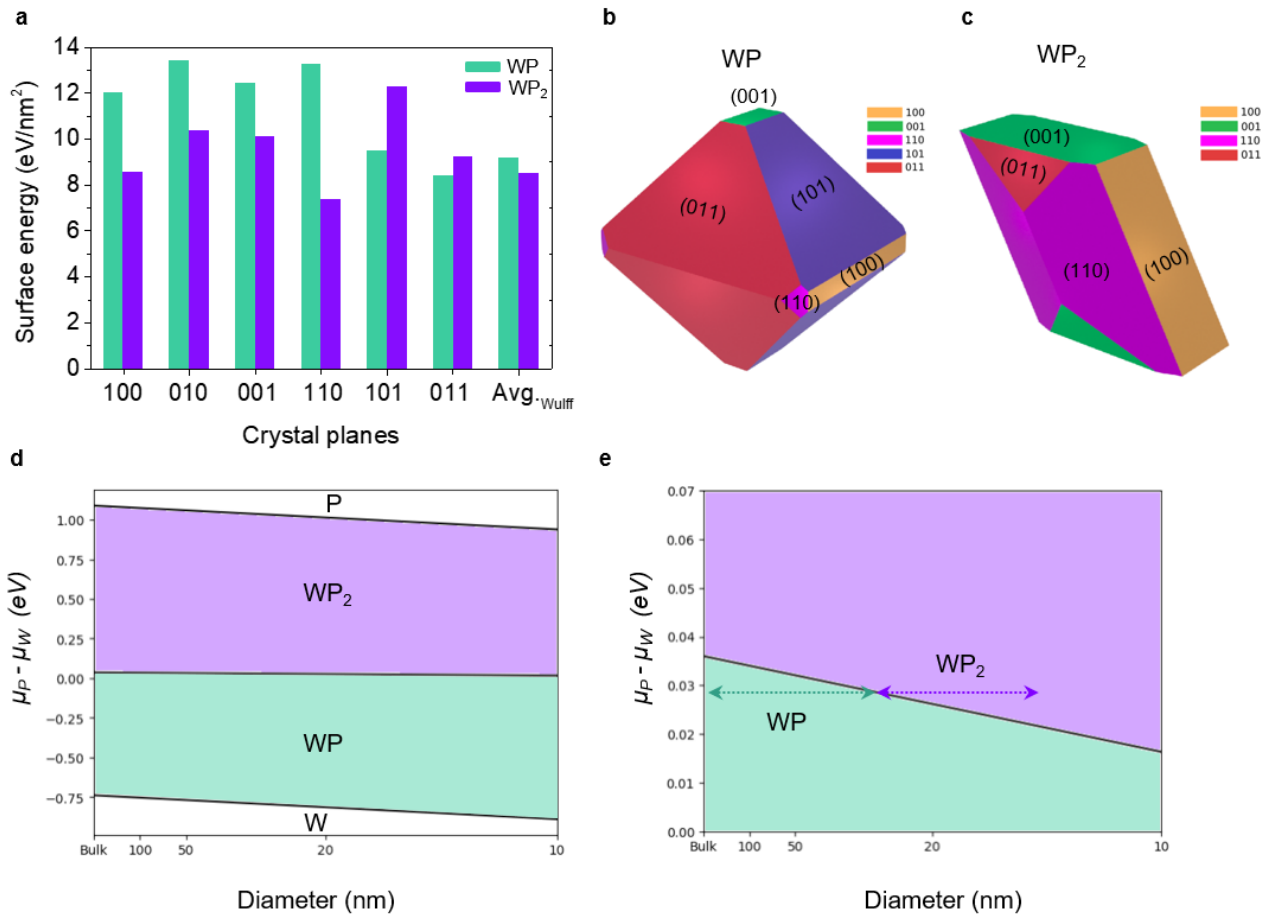

**Fig. 4| Size-dependent crystallization pathways of WP and α-WP₂.**
**a**, Calculated surface energies of various crystal planes for WP and α-WP₂. **b,c** Wulff polyhedron constructions for WP (**b**) and WP₂ (**c**). **d**, Phase diagram for tungsten and phosphorus as a function of diameter from bulk to 10 nm. **e**, Enlarged phase diagram for tungsten and phosphorus as a function of diameter. Arrows are estimated experimental conditions showing the transition of the phase at approximately 35 nm.

**Size-dependent phosphorization pathways of WP and α-WP₂**

Theoretical calculations were carried out to understand the experimentally observed diameter-dependent crystallization routes. We first calculated surface energies of various crystal planes for the two distinct phases of WP and α-WP₂ (Fig. 4a and Supplementary Table. 1), and then constructed appropriate Wulff shapes to minimize the total surface energy for each phase (Fig. 4b and 4c). We found that the average surface energy of the α-WP₂ Wulff shape (8.516 eV/nm$^2$) is lower than that of the WP Wulff shape (9.162 eV/nm$^2$), which suggests that α-WP₂ should be preferred over WP with decreasing diameter that corresponds to increasing surface-to-volume ratio. The surface energy comparison is consistent with our experimental observation that α-WP$_2$ is observed in nanowires with diameter below ~ 35 nm and WP above ~35 nm.

To quantify this phase transition, we compute the free energy, $g_{WP}$, of WP and WP$_2$ phases. In doing so, surface energy becomes a crucial parameter when assuming cylindrical nanowire geometry. We normalize this free energy per atom to allow for a direct comparison between materials, which provides the equation:

$$g_{WP} = E_{Formation_{WP}} - \frac{1}{2}\mu_{PW} + \frac{4E_{Surface}v_{WP}}{d}$$

where $E_{surface}$ is the average surface energy per unit area as calculated from the Wulff model, $v_{WP}$ is the unit cell volume per atom, $d$ is the nanowire diameter, and $\mu_{PW}=\mu_P - \mu_W$ is the chemical potential difference between P and W. In this equation, the factor of 1/2 corresponds to the mole fraction of P in WP. To obtain the corresponding equation for WP$_2$, this fraction would simply be modified to 2/3. For both phases, formation energy was taken from the Materials Project database.[41] A phase diagram was calculated as a function of $\mu_{PW}$ and $d$ by equating $g_{WP}$ and $g_{WP2}$ (Fig. 4d and 4e), which predicts a transition of stable phase from WP to α-WP$_2$ with decreasing diameter. We note that we had previously studied topological metal MoP nanowires using the same template-assisted growth method[11] and accordingly construct a diameter-dependent phase diagram for MoP (Supplementary Fig. 2), which shows a different behavior from the W-P phase diagram.

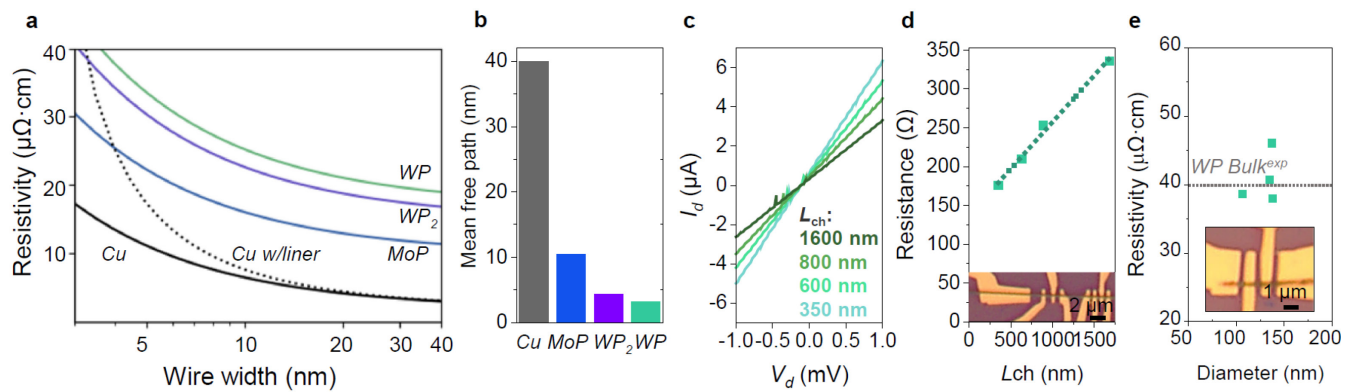

**Fig. 5| Resistivity scaling of WP.**

**a**, Calculated resistivity scaling of topological metal phosphides (WP, WP$_2$, MoP) as square wires, in comparison with Cu square wires with and without liner. Single crystal was assumed for the calculations with complete diffuse electron scattering at surfaces ($p = 0$). We note that experimental bulk resistivities ($r_0$) are used when width $\rightarrow \infty$. **b**, Calculated average mean free paths of Cu, MoP, WP$_2$, and WP. **c**, Channel-length ($L_{ch}$) dependent current-voltage ($I$-$V$) curves of the 1D-confined polycrystalline WP. **d**, $L_{ch}$ dependent resistance variation of the 1D-confined WP. Inset: Optical microscopy image of the measured device. **e**, Room temperature resistivity data of 1D-confined WP with varying cross-sectional area. Dotted line: resistivity value of WP bulk single crystal.[16]

**Electron transport properties of 1D-confined WP**

As discussed in the introduction, a subset of TMPs are topological metals that are promising as nanoscaled interconnects. To test the feasibility of these TMPs as interconnects beyond the current 3 nm technology nodes with the metal pitch of ~ 24 nm[42], room temperature resistance of these TMPs at dimensions below 35 nm must be obtained. Size-dependent room temperature resistivities of MoP, WP, and WP$_2$ were calculated for two representative geometries of square cross-section wires (Fig. 5a) and thin films (Supplementary Fig. 3) using Fuchs–Sondhemier models to assess electron scattering at surfaces (Calculation details in Supplementary Table. 2).[43,44] To minimize the surface scattering, the transport directions were chosen to be along the a- and c-axes for WP and WP$_2$, and the c-axis for MoP given their anisotropic Fermi surfaces (Supplementary Table. 3). For comparison, resistivities of Cu wires with and without a liner are also shown in Fig. 5a. For each resistivity curve, complete diffuse surface scattering was assumed with the specularity parameter $p = 0$.[45-47] We did not consider grain boundary scattering in these calculations because the calculated mean free paths of WP (3.15 nm) and WP$_2$ (4.33 nm) (Fig. 5b) were much smaller than the observed grain sizes (Fig. 3). The dimensional scaling of resistivity for WP and α-WP$_2$ appears similar to that of Cu without the liner and better than Cu with the liner. However, since the bulk resistivity of WP and WP$_2$ is higher than Cu, WP and WP$_2$ do not appear promising as low-resistance interconnects. Further, the resistivity scaling calculations suggests that for WP and WP$_2$, suppressed electron scattering originating from topologically protected surface states is negligible at room temperature, unlike the cases for CoSi and NbAs.[23,24]

We carried out room temperature resistivity measurements on the WP nanowires to experimentally gauge the degree of surface and grain boundary scattering as compared to the calculations. Using the transfer length methods, linear *I–V* curves were obtained by varying the channel length (Fig. 5c), where the channel resistance increased linearly with the channel length as expected and the contact resistance of 68.9 Ω was extracted (Fig. 5d). Four-probe resistance measurements were carried out on several WP nanowires (Fig. 5e). Remarkably, the resistivity values of our 1D-confined polycrystalline WP, which ranged between 38 and 46 μΩ·cm for cross-section areas of 8000 - 15000 nm$^2$, are comparable to that of bulk WP single crystal (40 μΩ·cm) grown by chemical vapor transport.[16] Supplementary Fig. 4 shows the correlation between the resistivity and grain structures for WP nanostructures with corresponding TEM images. Thus, in agreement with the calculations, the resistivity of 1D-WP implies that electron scattering at grain boundaries is negligible, which can be attributed to the small mean free path (3.15 nm).

**Conclusions**

We have demonstrated size-dependent phosphorization routes of 1D-confined WP and WP$_2$ from 1D WO$_2$ templates. Crystallographic orientations of grains and grain distributions for each phase were characterized using 4D-STEM, revealing critical transition regime of ~ 35 nm diameter below and above which WP$_2$ and WP form. The calculated surface energies from the Wulff constructions of these two competing phases predict α-WP$_2$ to be stable over WP at the nanoscale, which was further shown in the diameter-dependent phase diagrams obtained by DFT calculations. We achieved the lowest resistivity of 38 μΩ·cm for WP nanostructures, which is comparable to that of bulk single crystal, indicating electron scattering at grain boundaries and surfaces must be small despite the disordered polycrystalline nature of our samples likely due to their small mean free path. Our findings suggest that the diameter-dependent crystallization route can be exploited to guide synthesis protocols of 1D topological semimetals.

## Methods

### 1D-confined conversion of WP and WP$_2$

1D-WO$_2$ were grown by chemical vapor deposition. WO$_3$ source powder (Sigma-Aldrich, 99.95%) was placed at the center of the hot-walled tube furnace. Chamber pressure was maintained at 3 Torr with 20 sccm of H$_2$ gas. C-sapphire substrate with the miscut angle of 1° was placed at upstream of the furnace. The temperature of the upstream was maintained at 600 °C for 10 min. To convert the WO$_2$ templates to WP and WP$_2$, the as-grown 1D WO$_2$ templates were placed in the center of the furnace with 3 g of NaH$_2$PO$_2$·H$_2$O (Sigma-Aldrich, ≥99%) placed upstream. The chamber was pumped down to 100 mTorr, and then 30 sccm of H$_2$ was flowed until the furnace pressure reached atmospheric pressure. The temperature of the furnace was ramped up to 700 °C and held there for 50 minutes.

### 4D-STEM measurements

STEM experiments were performed on a C$_s$-probe-corrected Thermo Fisher Scientific Spectra 300 with an extreme-brightness cold field emission gun. The 4D-STEM measurements were collected using an EMPAD at 120 kV and a convergence angle of 0.5 mrad. The 4D-STEM datasets were processed using py4DSTEM and the associated crystal orientation mapping code[34,35]

### DFT Calculations

All first-principles calculations were performed using the open-source plane-wave software JDFTx.[47] Calculations for all materials were run using the Perdew-Burke-Ernzerhof (PBE) exchange-correlation functional[48] with a plane wave energy cutoff of 680 eV, and ultrasoft pseudopotentials were sourced from the GBRV library[49]. Self-consistent calculations were performed using a Gamma-cented mesh of 12x12x12 k-points to compute the bulk free energies of WP and WP$_2$, whereas surface-oriented slabs utilized a 12x12x1 k-point mesh. Slab geometries were constructed such that the slab thickness was approximately 30 angstroms, with a vacuum spacing of 15 Angstroms. A structural relaxation was performed iteratively for bulk and slab structures to optimize the lattice constants and atomic positions. Upon calculating surface energies for WP and WP$_2$, Wulff models were constructed using the WulffPack Python package[50]. For the calculation of bulk resistivities, all electronic structure and phonon properties were transformed into the maximally localized Wannier function basis[51]. In the case of WP, a total of 48 Gaussian Wannier centers were iteratively fitted to the band structure in the energy range -15.3 eV to +4.55 eV, relative to the VBM, and a phonon q-mesh of 4x2x2 was chosen. For WP$_2$, a total of 48 Gaussian Wannier centers were fitted in the energy range -16.9 eV to +4.38 eV, and a phonon q-mesh of 2x4x2 was chosen.

## Device fabrication of 1D-confined WP

The converted WP crystals were transferred onto SiO$_2$/Si substrates using a PMMA-assisted wet-transfer method, then coated with e-beam resist layers (200 nm MMA EL 8.5 and 200 nm PMMA A3). Electrode patterns for transfer-length methods and four-probe measurements were written by standard e-beam lithography using a ThermoFisher Helios G4 system. 10/100 nm-thick Cr/Au electrical contacts were deposited by e-beam evaporation.

## Data availability

All data that support the findings of this study are available within the paper and its Supplementary information.

## Acknowledgements

Synthesis and 4D STEM experiments of tungsten phosphides were supported by the Gordon & Betty Moore EPiQS Initiative, grant GBMF9062. Theoretical calculations and transport measurements were supported by the SRC nCORE IMPACT under Task 2966.002 and 2966.005.

Author information
These authors contributed equally: Gangtae Jin, Christian D. Multunas, James L. Hart.
Authors and Affiliations

**Department of Materials Science and Engineering, Cornell University, Ithaca, New York 14850, USA**

Gangtae Jin, James L. Hart, Mehrdad T. Kiani, Quynh P. Sam, Han Wang, Yeryun Cheon, Khoan Duong, Judy J. Cha

**Department of Materials Science and Engineering, Rensselaer Polytechnic Institute, Troy, NY, 12180 USA**

Christian D. Multunas, Ravishankar Sundararaman

Department of Mechanical Engineering and Materials Science, Yale University, New Haven, Connecticut 06511, USA

David J. Hynek

Department of Environment and Energy Engineering, Sungshin Women's University, Seoul 01133, Republic of Korea

Hyeuk Jin Han

Contributions

J.J.C. and G.J. supervised the project. G.J. performed synthesis, material characterization and electrical measurements. C.D.M. and R.S. performed DFT calculations. J.L.H., H.J.H. and Q.P.S. performed TEM experiments and data analysis. M.T.K., H.W., Y.C., K.H., D.J.H. carried out further materials characterization. G.J, C.D.M, J.L.H., H.J.H., R.S. and J.J.C. co-wrote the manuscript. All authors have read the manuscript and commented on it.

**Ethics declarations**

Competing interests

The authors declare no competing interests.

**Additional Information**

Supporting Information. The online version contains supplementary material available at

Correspondence and requests for materials should be addressed to judy.cha@cornell.edu, sundar@rpi.edu, hyeukjin.han@sungshin.ac.kr.

**Figure Captions**

**Fig. 1| Electron band structures and template-assisted conversion of WP and WP$_2$.**

**a,b,** Crystal structure (**a**) and calculated surface-weighted electronic band structure (**b**) of a WP slab, terminated with the lowest surface energy plane (011). Surface states from the terminated crystal planes are colored in red. Note that these surface states do not represent topologically protected surface states. **c,d,** Crystal structure (**c**) and calculated surface-weighted electronic band structure (**d**) of a α-WP$_2$ slab, terminated with the lowest surface energy plane (110). Surface states from the terminated crystal planes are colored in red. We have chosen the same high-symmetry path for orthorhombic WP and monoclinic WP$_2$ since these band structures represent slabs rather than bulk structures, and the slabs have identical surface unit cells. **e,** Schematics of geometrically confined transformation from WO$_2$ to WP with corresponding crystal structures of WO$_2$ and WP. **f,** Optical microscope images of WO$_2$ template grown on c-cut sapphire and transformed 1D-WP. scale bars: 10 μm. **g,** Powder X-ray diffraction spectra of growth products as a function of the conversion temperature. **h,** Relative atomic content of W$_x$P$_{1-x}$ (0.5<x<0.7) growth products, collected from SEM- EDS signals at 4 different conversion temperatures. **i,** Raman spectra of as grown 1DWO$_2$ (orange), transformed 1D-WP (green), and WO$_3$ (blue), respectively. a.u., arbitrary units.

**Fig. 2| Structural characterization of 1D-confined WP$_{x\ (x\ =\ 1\ or\ 2)}$ with varying diameter.**

**a,** Low-magnification TEM image of 1D-confined WP. **b,** Atomic-resolution HAADF-STEM images of WP obtained from a single grain of the polycrystalline WP. **c,d,** HAADF-STEM images of 1D-confined WP with the diffraction pattern of dominant zone axis of [131] (**c**) and WP$_2$ with the corresponding diffraction pattern of zone axis of [233] (**d**). **e,** TEM-EDS spectra acquired from representative 1D-confined WP$_{x\ (x\ =\ 1\ or\ 2)}$. **f,** Relative atomic content of P with varying diameter, collected from TEM-EDS signals of 5 different samples.

**Fig. 3| 4D STEM and grain orientation mapping of WP and α-WP$_2$.**

**a,** Schematic for 4D-STEM of 1D-confined WP$_{x\ (x\ =\ 1\ or\ 2)}$ on EMPAD. **b,c** HAADF-STEM image (**b**) and grain orientation map (**c**) of 1D-confined orthorhombic WP. The orientation map only shows a sub-section of the wire. **d,** Inverse pole figure for WP and electron diffraction patterns with simulated diffraction patterns from specific grains of orthorhombic WP. **e,f** HAADF-STEM image (**e**) and grain orientation map (**f**) of 1D-confined monoclinic WP$_2$. **g**, Inverse pole figure for WP$_2$ and electron diffraction patterns with simulated diffraction patterns from specific grains of monoclinic WP$_2$. Scale bars: 100 nm for **b** and **e**, 50 nm for **c** and **f**. For the overlapping experimental and simulated diffraction patterns, the label provides the nearest zone axis.

**Fig. 4| Size-dependent crystallization pathways of WP and α-WP$_2$.**

**a**, Calculated surface energies of various crystal planes for WP and α-WP$_2$. **b,c** Wulff polyhedron constructions for WP (**b**) and WP$_2$ (**c**). **d**, Phase diagram for tungsten and phosphorus as a function of diameter from bulk to 10 nm. **e**, Enlarged phase diagram for tungsten and phosphorus as a function of diameter. Arrows are estimated experimental conditions showing the transition of the phase at approximately 35 nm.

**Fig. 5| Resistivity scaling of WP.**

**a**, Calculated resistivity scaling of topological metal phosphides (WP, WP$_2$, MoP) as square wires, in comparison with Cu square wires with and without liner. Single crystal was assumed for the calculations with complete diffuse electron scattering at surfaces ($p = 0$). We note that experimental bulk resistivities ($r_0$) are used when width → ∞. **b**, Calculated average mean free paths of Cu, MoP, WP$_2$, and WP. **c**, Channel-length ($L_{ch}$) dependent current-voltage ($I$-$V$) curves of the 1D-confined polycrystalline WP. **d**, $L_{ch}$ dependent resistance variation of the 1D-confined WP. Inset: Optical microscopy image of the measured device. **e**, Room temperature resistivity data of 1D-confined WP with varying cross-sectional area. Dotted line: resistivity value of WP bulk single crystal.[16]

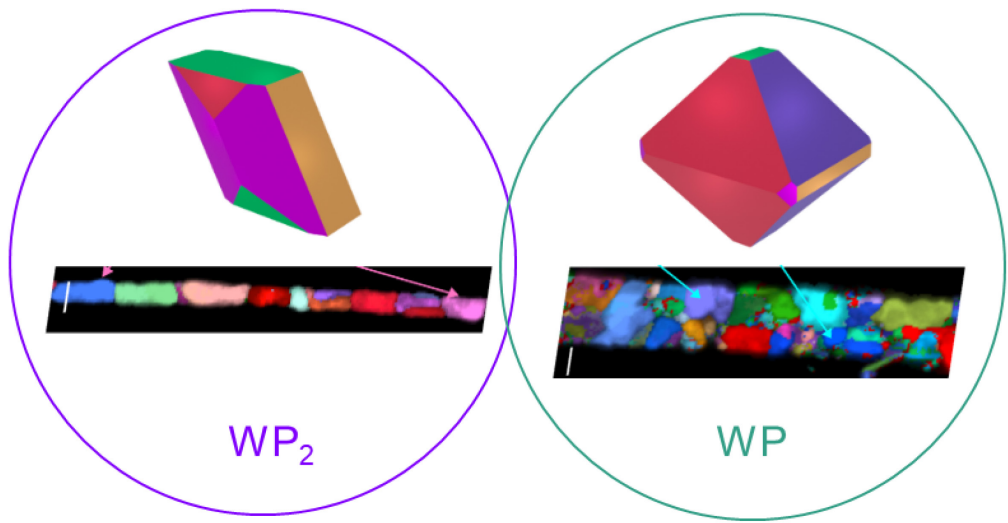

**Table of Contents artwork**